\documentclass[11pt]{article}
\usepackage[a4paper]{geometry}
\usepackage{amsthm}
\usepackage{amsmath}
\usepackage{amssymb}
\usepackage{amsfonts}
\usepackage{epsfig}
\usepackage[bf,SL,BF]{subfigure}
\usepackage[all]{xypic}

\begin{document}

\theoremstyle{theorem}
\newtheorem{Lemma}{Lemma}[section]
\newtheorem{Proposition}{Proposition}[section]
\newtheorem{Corollary}{Corollary}[section]
\newtheorem{Theorem}{Theorem}[section]
\newtheorem{Condition}{Condition}[section]
\newtheorem{Assumption}{Assumption}[section]
\newtheorem{Integrator}{Integrator}

\theoremstyle{definition}
\newtheorem{Definition}{Definition}[section]
\newtheorem{Remark}{Remark}[section]
\newtheorem{Example}{Example}[section]

\title{Temperature and Friction Accelerated Sampling of Boltzmann-Gibbs Distribution}

\author{Molei Tao, Houman Owhadi, Jerrold E. Marsden}

\maketitle

\begin{abstract}
This paper is concerned with tuning friction and temperature in Langevin dynamics for fast sampling from the canonical ensemble.
We show that near-optimal acceleration is achieved by choosing friction so that the local quadratic approximation of the Hamiltonian is a critical damped oscillator. The system is also over-heated and cooled down to its final temperature. The performances of different cooling schedules are analyzed as functions of total simulation time.
\end{abstract}

\section{Introduction}
We propose a method to accelerate the Langevin approach of sampling from Boltzmann-Gibbs (B-G) distribution. Specifically, consider the following Langevin Stochastic Differential Equations (SDE)
\begin{equation}\label{sjhdgjhsggde}
    \left\{ \begin {array} {rcl}
        M dq &=& pdt \\
        dp &=& -\nabla V(q)dt-cpdt+\sqrt{2c/\beta} dW
    \end {array} \right.
\end{equation}
where $p,q\in \mathbb{R}^d$, $M$ is the mass matrix, $V(\cdot)$ is potential energy, $c$ is a positive semi-definite $d\times d$ matrix indicating the damping coefficient, $\beta\in\mathbb{R}^+$ is the inverse of temperature, and $W$ is a standard Wiener process.

It is known that the stochastic process defined by \eqref{sjhdgjhsggde} has an invariant distribution of Boltzmann-Gibbs distribution (also known as canonical ensemble) defined by:
\begin{equation}
    d\mu = Z^{-1}\exp(-\beta H(q,p))dqdp.
\end{equation}
where $Z=\int_{T^*\mathbb{R}^d} \exp(-\beta H(q,p))dqdp$ is the partition function, and $H(q,p)=p^T M^{-1} p/2+V(q)$ is the Hamiltonian function.

When the solution of \eqref{sjhdgjhsggde}  is also geometrically ergodic with respect to $\mu$
(we refer to \cite{MR1924935} and \cite{MaStHi02} for sufficient conditions on the potential $V$), it is then natural to use long-time trajectories of \eqref{sjhdgjhsggde} as approximate samples of B-G distribution.

One important thing to notice is that being able to sample from B-G enables sampling an arbitrary probability density function. The trick is to set $V(q)=-\beta^{-1}\ln \pi(q)$, and then the marginal distribution on $q$ from B-G will have the density function $\pi(\cdot)$.

This paper is concerned with the following questions:
\begin{itemize}
\item Although the friction parameter $c$ does not affect the invariant distribution, it does affect the rate of convergence. How should $c$ be chosen for faster convergence and hence accelerated sampling?
\item If sampling from B-G is the objective, the inverse temperature $\beta$ does not need to be kept constant over the total simulation time $T$.
How to chose the cooling schedule $t\mapsto \beta(t), t\in [0,T]$ in order to minimize the distance between the distribution of $[q(T),p(T)]$ and the desired B-G?
\end{itemize}

\paragraph{Background:}
There is no need to repeat the importance of sampling the canonical ensembles of complicated systems, which is, however, a known computational challenge \cite{BaLaLe03, HaMoMc04, KaLa93}. The nonlinearity of the potential and the curse of dimensionality, for instance, make sampling methods slowly convergent.

Classical sampling approaches include purely statistical methods such as Metropolis algorithm and importance sampling that are solely for sampling purposes (see for instance \cite{BaDi05} and references therein for a review and comparison), stochastic molecular dynamics (primarily Langevin dynamics), deterministic dynamics plus an external thermostat (such as Nos\'{e}-Hoover \cite{Nose84, Hoover85}, Berendsen \cite{Be84} or Andersen \cite{An80} thermostats), Hybrid Monte Carlo \cite{DuKe87} (which introduces auxiliary dynamics to avoid random walks), etc. We also refer to \cite{RoTw96} as an example that combines stochastic molecular dynamics and purely statistical approach.

 Langevin dynamics adds friction and noise to mechanical equations to model energy exchange with a heat bath \cite{TuLaFr77, Sc02, ScSt78}.  It has been shown in the context of classical molecular sampling that both stochastic dynamics and  deterministic dynamics with thermostats outperform purely statistical methods in  convergence rate as the size of the system grows (we refer to \cite{CaLeGa07} for a linear alkane molecule). Since overdamped Langevin is a special case of Hybrid Monte Carlo \cite{ChBe09}, it is not surprising to observe cases in which Langevin dynamics is computationally more efficient than purely statistical methods. Moreover, if the system is stiff or multiscale, existing stiff or multiscale Langevin integrators such as SIM \cite{SIM1} or FLAVOR \cite{FLAVOR09} can be directly employed for accelerated computation.

 Annealing was first introduced in Simulated Annealing algorithm \cite{KiGeVe83} for global optimization, which can also be viewed as (uniformly) sampling from the set of global minimizers of $V$. Temperature accelerated dynamics has been proposed in
 \cite{TAD00} for   events simulations. The concept there is to raise temperature of the  system to make rare events occur more frequently, intercept each attempted escape from potential wells and extrapolate time to low temperature. Another temperature approach has been used to calculate free energy \cite{MaVa06}. In that method, overheated auxiliary variables are introduced to equilibrate the collective variables faster. We stay with the global annealing approach used in Simulated Annealing.

The proposed perspective of tuning friction and annealing temperature is distinct from prevailing accelerated sampling methods, such as conformational flooding \cite{Gr95}, replica exchange \cite{SuOk99}, umbrella sampling \cite{ToVa77}, self-guided MD \cite{WuWa99}, hyperdynamics \cite{Vo97}, affine invariant ensemble sampler \cite{GoWe09}, and many others reviewed in \cite{BeSt97}, and therefore can be used concurrently with many of these methods. While tuning friction is mostly restricted to dynamics based methods, annealing may apply to any method that involves temperature. Note that temperature is a rather general notion because it can often be introduced artificially; for instance, see \cite{Ne96} for an example in which temperature is introduced in an MCMC algorithm for Bayesian updating.

\section{Method for friction and temperature accelerated Boltzmann-Gibbs sampling}

\paragraph{Background algorithms:}
Although any Langevin integrator can serve as a background algorithm and be tuned and annealed, in our numerical simulations we base on the 1st-order B-G preserving Geometric Langevin Algorithm (GLA) introduced in \cite{BoOw:09}, which is recapped as follows:

\begin{equation}
\left\{ \begin {array} {rcl}
    \hat{p}_n &=& e^{-c_n h}p_n + \sqrt{\frac{1-e^{-2c_n h}}{\beta_n}} \xi_n \\
    q_{n+1} &=& q_n+h\hat{p}_n \\
    p_{n+1} &=& \hat{p}_n-h\nabla V(q_{n+1})
\end {array} \right.
\label{GLA}
\end{equation}
where $h$ is the timestep length, $\xi_n$'s are i.i.d. standard normal random variables, and $c_n=c$ and $\beta_n=\beta$ in absence of friction tuning or temperature annealing.

The choice of GLA is motivated by its conformal-symplecticity and long-time properties \cite{BoOw:09}. Specifically, under certain conditions, GLA is not only pathwise accurate but also convergent towards B-G up to a diminishing numerical error. It is worth mentioning that similar properties are shown to hold under weaker conditions for a Metropolized version of GLA \cite{RoTw96, BoVa:09}, which can also be tuned and annealed for accelerated samplings.

For multiscale or stiff systems (where $V(q)=V_0(q)+\epsilon^{-1}V_1(q)$ for instance), FLAVORS \cite{FLAVOR09} are possible alternative background algorithms that are also conformal-symplectic (we also refer to SIMS \cite{SIM1} for quadratic stiff potentials).

\paragraph{Choice of friction:}
If $V$ is quadratic (of the form $V=\frac{q^T K q}{2}$), we show in Appendix \ref{linearSection} that optimal acceleration is achieved by choosing $c=2 K^\frac{1}{2}$ so that all degrees of freedom of the harmonic oscillator are critically damped. Based on this observation, we heuristically propose to tune the friction $c_n$ at each time step of the simulation according to the Hessian of the potential $V$:
\begin{equation}
    \begin{cases}
        k_n &= \begin{cases}
                        \frac{1}{2} \frac{\partial ^2 V}{\partial q^2}(q_n) , & \frac{\partial ^2 V}{\partial q^2}(q_n) \succ 0 \\
                        \alpha^2/4 I, & \text{otherwise}
                \end{cases} \\
        c_n &= 2\sqrt{k_n}
    \end{cases}
\end{equation}
where $\alpha$ is a fixed real parameter, preassigned to handle the case of negative curvature; for instance, it could be equal to $0$ or to the original value of $c$.

\paragraph{Choice of temperature:}
Annealing has successfully been applied to optimization problems \cite{KiGeVe83}.  A cooling schedule describes how to choose $T(n)=1/\beta_n$ as a function of $n$. For optimization based cooling schedules, one requires $\lim_{i\rightarrow \infty} T(i)=0$. We refer to \cite{Ha02,CoFi99,TrCoSi05} for general reviews of optimization based cooling schedules, and to \cite{GeGe84,Ha88} for theoretical bounds on convergence.
In this paper we are interested in situations where the total number of steps $N$ is finite and fixed, the final temperature $T(N)=T_f=1/\beta>0$ is strictly positive and is the temperature at which one wishes to sample the B-G distribution.

It is then natural to seek to minimize the distance between the distribution of $(q_N,p_N)$ and B-G at temperature $1/\beta$ using $T(1),\ldots,T(N-1)$ as optimization variables. In Appendix \ref{Sectemaccsam1} we derive a bound on this distance using transition state theory and convergence rates of Markov chains. A numerical minimization of that bound suggests the following near-optimal cooling schedule for $T_f>0$ (for $T_f=0$ we refer  to \cite{CoFi99} and references therein) and $N<N_0$ ($N_0$ is the number of steps needed for sampling by a naive Langevin simulation; see Appendix \ref{Sectemaccsam2} for details):
\begin{equation}
    \beta_n= \frac{n}{N}\frac{1}{T_f}+(1-\frac{n}{N})\frac{1}{T_i}, \quad T(n)=1/\beta_n
\end{equation}
where $N$ is the total-number of simulation steps, $T_f$ the temperature at which the Gibbs distribution needs to be sampled, and the initial temperature $T_i>T_f$ is a free parameter chosen to overcome the maximal potential barrier, i.e., $T_i\gg \Delta V/k$ (for simplicity we let the Boltzmann constant $k$ be equal to one in our setting; $\Delta V$ can be intuitively interpreted as the maximum elevation in potential landscape, and we refer to \cite{DiSt91} for a rigorous definition).

\paragraph{Friction and temperature accelerated sampling:}
Put together, annealed and tuned GLA (AnnealTuneGLA) for accelerated B-G sampling is the following:
\begin{equation}
\left\{ \begin {array} {rcl}
    k_n &=& \left\{ \begin {array} {ll}
                    \frac{1}{2} \frac{\partial ^2 V}{\partial q^2}(q_n) & \frac{\partial ^2 V}{\partial q^2}(q_n) \succ 0 \\
                    \alpha^2/4 & otherwise
                \end{array} \right. \\
    c_n &=& 2\sqrt{k_n} \\
    \beta_n &=& \frac{n}{N}\frac{1}{T_f}+(1-\frac{n}{N})\frac{1}{T_i} \\
    \hat{p}_n &=& e^{-c_n h}p_n + \sqrt{\frac{1-e^{-2c_n h}}{\beta_n}} \xi_n \\
    q_{n+1} &=& q_n+h\hat{p}_n \\
    p_{n+1} &=& \hat{p}_n-h\nabla V(q_{n+1})
\end {array} \right.
\label{TuneGLA}
\end{equation}
Comparing to the background GLA, the distribution of the accelerated trajectory at a fixed time is closer to the desired B-G in the total variation sense. A possible exact preservation of a near-by distribution is, however, not yet proved for AnnealTuneGLA.

It is worth mentioning that 1st-order GLA is not unconditionally stable, nor is AnnealTuneGLA. Therefore, $h$ or $\alpha$ should not be chosen to be too large.

\section{Numerical experiments}
\label{exampleSection}

\begin{figure}[h]
\centering
\includegraphics[width=0.5\textwidth]{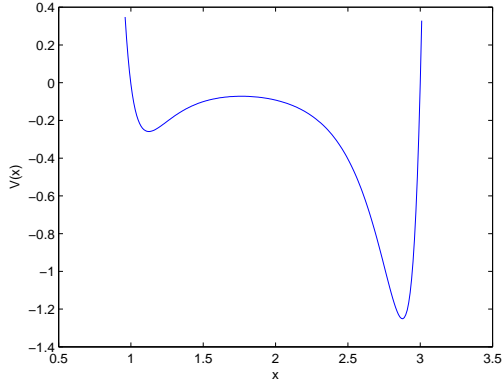}
\caption{\footnotesize Potential energy landscape.}
\label{potentialLandscape}
\end{figure}

Consider a one dimensional nonlinear molecular system consisting of two distinct  heavy (fixed) atoms and a light atom between them. It is modeled as a single degree of freedom Hamiltonian system with a Lennard-Jones potential function $V(q)=\left(q^{-12}-q^{-6}\right)+5\left((4-q)^{-12}-(4-q)^{-6}\right)$ (Figure \ref{potentialLandscape}). The energy landscape consists of a local potential barrier and two potential wells. The attraction due to the right atom is larger than the left one. If one starts the dynamics with zero initial momentum and position in the left basin, the asymptotic (long time) position distribution will be a marginal of B-G and concentrated in the right basin. Therefore, the expectation of position $q$ at a fixed time can be used as an indicator of the convergence rate for this nonlinear system.

\begin{figure}[h]
\centering
\includegraphics[width=\textwidth]{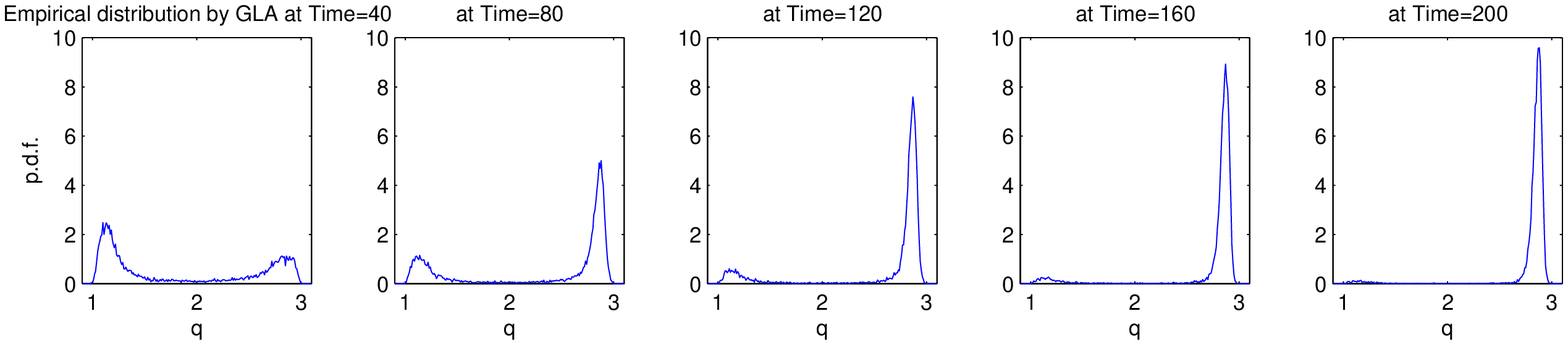}
\caption{\footnotesize Evolution of the empirical distribution obtained by GLA (Eq. \ref{GLA}) with $c=0.1$. The Markov process is converging as the distribution peaks more and more in the right potential basin. Simulation is done with a step length $h=0.01$ and distributions are approximated empirically by an ensemble of 10000 trajectories.}
\label{naiveGLA}
\end{figure}

\begin{figure}[h]
\centering
\includegraphics[width=0.5\textwidth]{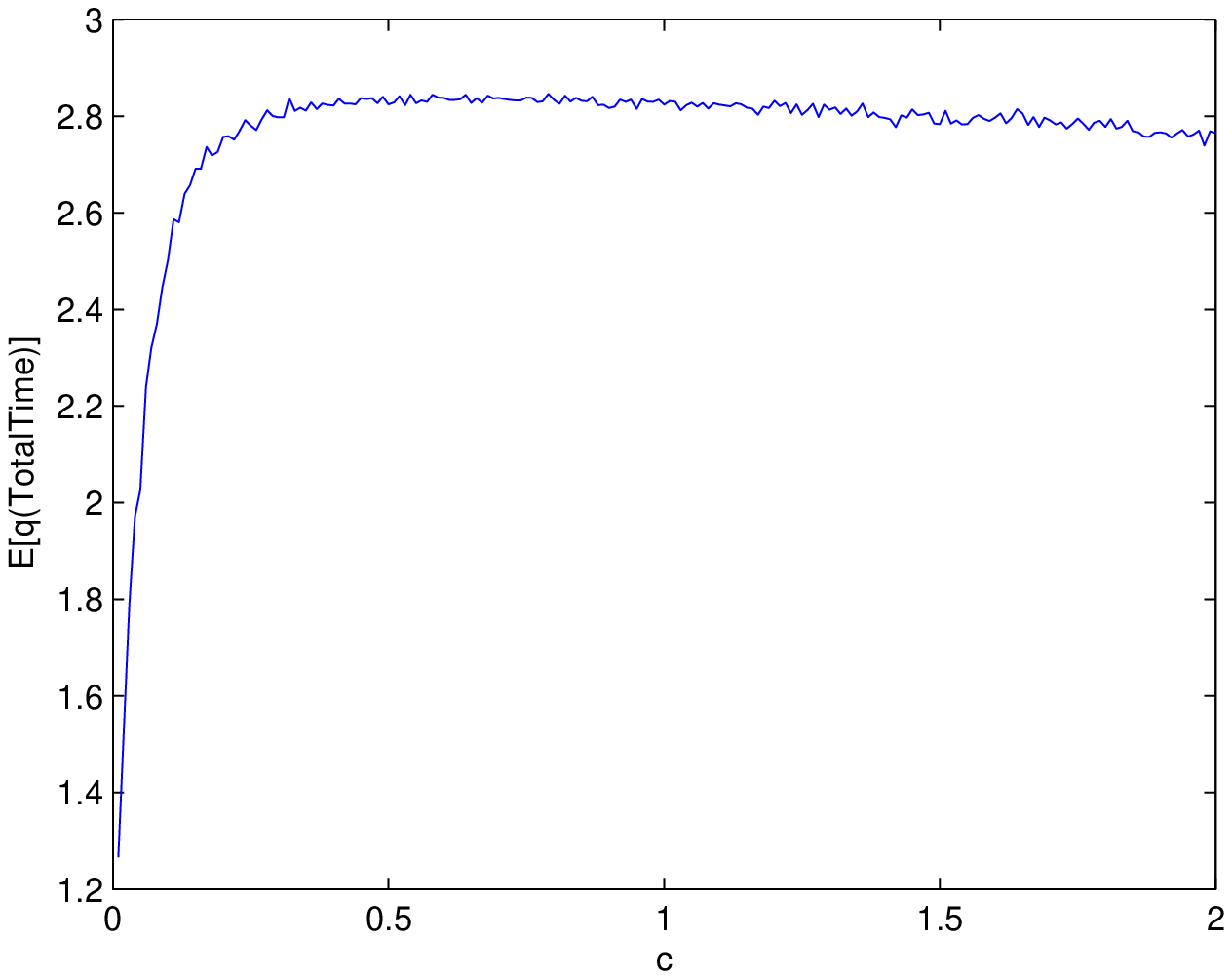}
\hspace{-20pt}
\includegraphics[width=0.5\textwidth]{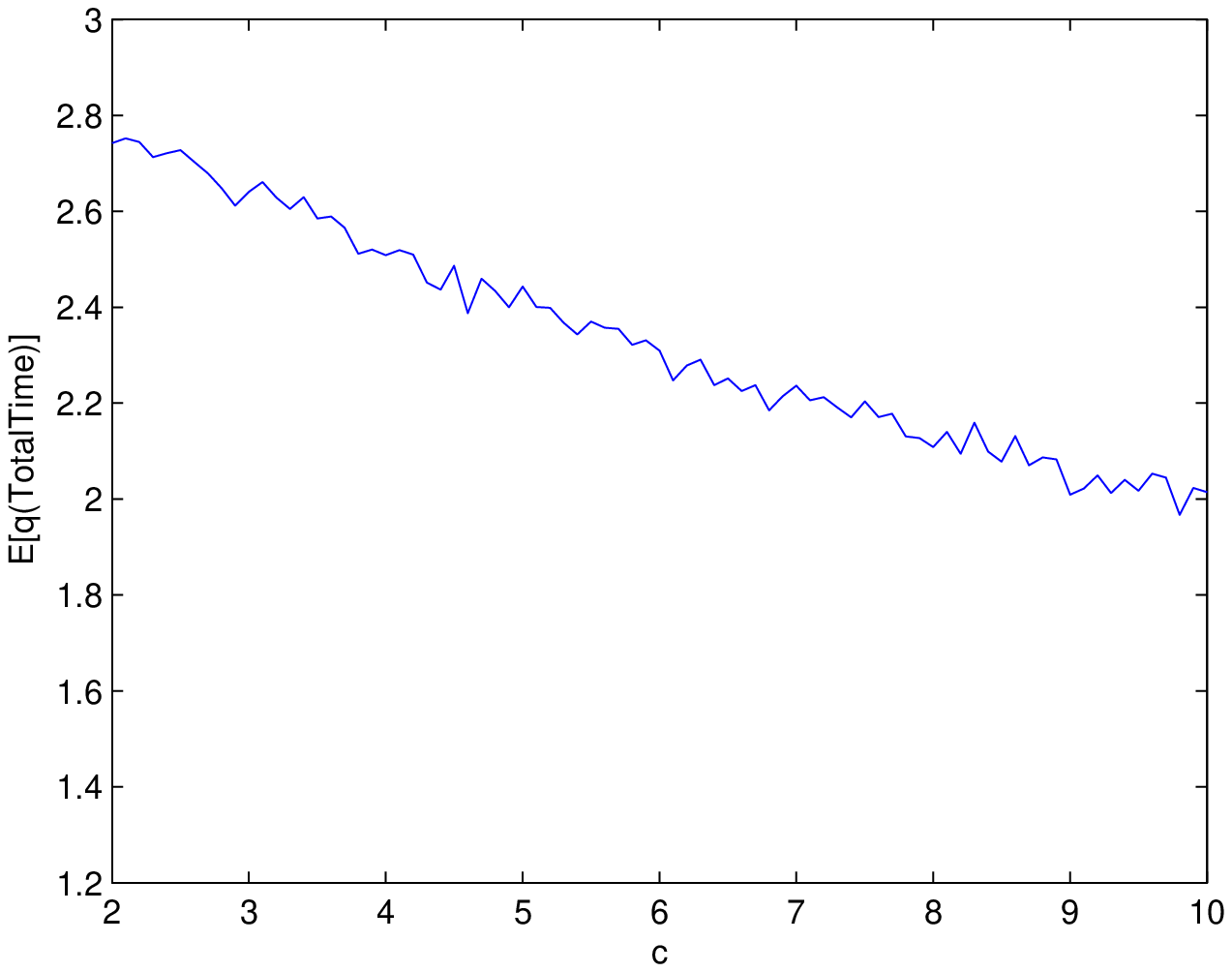}
\caption{\footnotesize Expectations of position at a fixed time for different frictions obtained by GLA (Eq. \ref{GLA}). Larger expectation implies better convergence in this problem, and therefore this indicates the relationship between choice of $c$ and convergence rate. The fixed time is TotalTime=100, step length is $h=0.01$, expectations are calculated by an empirical average over an ensemble of 1000 trajectories. $c$ values are enumerated from 0.01, 0.02, $\ldots$, 1.99, 2.00 and 2.10, $\ldots$, 19.90, 20.00.}
\label{GLAoptimization}
\end{figure}

Throughout this section we use parameters $\beta=10$, $q(0)=1.1$ and $p(0)=0$. With an arbitrarily chosen $c=0.1$, Langevin dynamics integrated with a B-G preserving method GLA (Eq. \ref{GLA}) takes more than $200$ time units before indiscernible convergence (Figure \ref{naiveGLA}). Enumerating $c$ values for fixed $\beta$ (and hence temperature $T$), one obtains different values of $\mathbb{E}[q(\text{TotalTime})]$ for a fixed total simulation time (Figure \ref{GLAoptimization}). This confirms that  the value of $c$  affects the convergence rate. The optimal fixed value is $c=0.7$ in this example.

\begin{figure}[h]
\centering
\includegraphics[width=\textwidth]{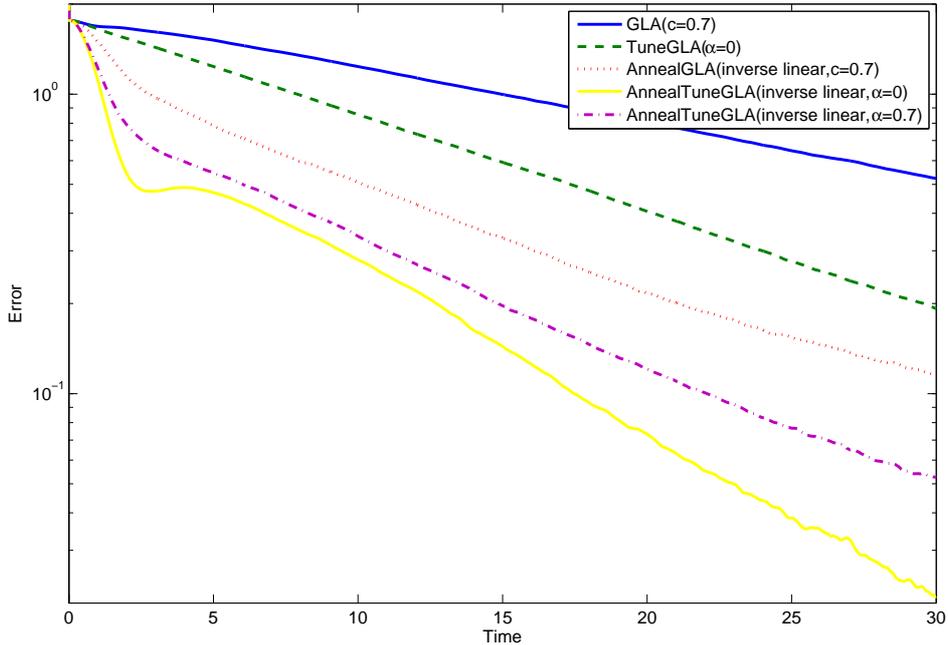}
\caption{\footnotesize Comparison of errors of GLA, TuneGLA with $c$ adaptively tuned, AnnealGLA with inverse linear cooling schedule, and AnnealTuneGLA with both. $c=0.7$ that ensures fastest GLA convergence (Figure \ref{GLAoptimization}) is used in GLA and AnnealGLA. A comparison between choices of $\alpha$ (which indicates the value of $c$ when curvature of potential is negative for tuning (Eq. \ref{TuneGLA}) is also presented. Total simulation time=30 is fixed, and error at each step throughout the simulation is recorded. Simulation step length is $h=0.01$. Error at time $t$ is calculated by $| \frac{1}{M}\sum_{i=1}^{M} q^i(t)-\mathbb{E}q(\infty) |$, where $M=10000$ is the total number of independent trajectories, $q^i(t)$ is the position of the $i$th trajectory at time $t$, and $\mathbb{E}q(\infty)$ is well approximated by empirical average of an ensemble of $20000$ GLA trajectories at total simulation time of $300$. The constant of initial temperature in the inverse linear cooling (Eq. \ref{AnnealGLA}) is $C=10T_f$ and applies to all three AnnealGLAs.}
\label{GLAcomparison}
\end{figure}

Although in practice it is rarely the case that an optimization can be carried out beforehand to determine the best value of $c$ for fastest convergence of GLA, we nevertheless use GLA with the optimal friction $c=0.7$ for comparison purposes. We will show that TuneGLA outperforms even this optimized GLA, demonstrating that $c$ really needs to be tuned locally.

In Figure \ref{GLAcomparison}, GLA with $c=0.7$ (the optimal fixed value), TuneGLA (GLA with friction tuning) which adaptively tunes $c$ but does not anneal (Eq. \ref{TuneGLA}), AnnealGLA (GLA with temperature annealing) which uses an inverse linear cooling schedule ($C=10T_f$) but does not tune $c$ (Eq. \ref{AnnealGLA}), and AnnealTuneGLAs that tune and anneal with respectively $\alpha=0$ and $\alpha=0.7$ are compared. We observe that tuning friction and annealing temperature individually accelerates the convergence, and their effects are additive. Therefore, the proposed AnnealTuneGLA has the fastest rate of convergence. In addition, here the choice of $\alpha=0$ slightly outperforms $\alpha=0.7$, which is set to be the value of the optimal $c$. The optimal choice of $\alpha$ has not been investigated.

\section{Acknowledgement}
This work is supported by NSF grant CMMI-092600. We are grateful to James L. Beck and Konstantin Zuev for insightful discussions.

\bibliographystyle{prsty}
\bibliography{molei14}

\section{Appendix}

\subsection{Friction accelerated sampling: analysis of linear systems}
\label{linearSection}
In this section, we will show that with $\beta$ fixed, the choice of $c=2\sqrt{k}$ will enable the fastest convergence of the following system:
\begin{equation}
    \left\{ \begin {array} {rcl}
        dq &=& pdt \\
        dp &=& -kqdt-cpdt+\sigma dW
    \end {array} \right.
    \label{linearSystem}
\end{equation}
where $\sigma=\sqrt{2c/\beta}$. Assume $k$ is a scalar for the moment. For our purpose, consider positive $k$, because if $k$ is 0 the system decouples, and if $k$ is negative the system is not ergodic and does not admit an invariant distribution.

The solution to the above linear system can be explicitly written as
\begin{equation}
    \left\{ \begin {array} {rcl}
        q(t) &=& B_{11}(t)q(0)+B_{12}(t)p(0)+\int_0^t B_{12}(t-s) \sigma dW_s \\
        p(t) &=& B_{21}(t)q(0)+B_{22}(t)p(0)+\int_0^t B_{22}(t-s) \sigma dW_s
    \end {array} \right.
\end{equation}
where $B(t)$ is the fundamental matrix defined through the following autonomous ODE $\frac{dB}{dt}=\begin{bmatrix}0&1\\-k&-c\end{bmatrix}B$, and written in block form to be
\begin{eqnarray}
    B(t)=\begin{bmatrix}B_{11}(t)&B_{12}(t)\\B_{21}(t)&B_{22}(t)\end{bmatrix}=\exp\left(\begin{bmatrix}0&1\\-k&-c\end{bmatrix} t \right)
    \label{FundamentalMatrix}
\end{eqnarray}

After calculating out the matrix exponential, the expectation of position writes as follows
\begin{eqnarray}
    \mathbb{E}q(t) &=& B_{11}(t) q(0) + B_{12}(t) p(0) \nonumber\\
    &=& \frac{e^{\frac{1}{2} \left(-c+\sqrt{c^2-4 k}\right) t} \left(c+\sqrt{c^2-4 k}\right)-e^{\frac{1}{2} \left(-c-\sqrt{c^2-4 k}\right) t} \left(c-\sqrt{c^2-4 k}\right)}{2 \sqrt{c^2-4 k}} q(0) \nonumber\\
    &+& \frac{e^{\frac{1}{2} \left(-c+\sqrt{c^2-4 k}\right) t}-e^{\frac{1}{2} \left(-c-\sqrt{c^2-4 k}\right) t}}{\sqrt{c^2-4 k}} p(0)
\end{eqnarray}

Naturally, the expectation approaches $0$ as $t\rightarrow +\infty$. Recall that $c$ and $k$ are nonnegative reals. We will show in the following discussion that the maximum speed of convergence toward $0$ will be achieved when $c=2\sqrt{k}$:
\begin{enumerate}
\item
When $c^2-4k>0$, $-c-\sqrt{c^2-4k}<-c+\sqrt{c^2-4k}<0$ and none of the coefficients are zero. Therefore the bottleneck for convergence of $B_{11}(t)$ and $B_{12}(t)$ will be $e^{\frac{1}{2} \left(-c+\sqrt{c^2-4 k}\right) t}$, which will be minimized as $c^2 \downarrow 4k$.
\item
When $c^2-4k=0$, $B_{11}=\frac{1}{2}e^{-ct/2}(2+ct)$ and $B_{12}=e^{-ct/2}t$.
\item
When $c^2-4k<0$, define a real number $\omega=\sqrt{4k-c^2}$. $B_{11}=e^{-ct/2}(c\sin(\omega t/2)/\omega+\cos(\omega t/2))$ and $B_{12}=e^{-ct/2}2\sin(\omega t/2)/\omega$. Notice $\cos(\omega t/2)$ and $\sin(\omega t/2)$ can not be simultaneously zero, and therefore the convergence rate is controlled by $e^{-ct/2}$, which will be minimized when $c^2 \uparrow 4k$.
\end{enumerate}

Hence when $c=2\sqrt{k}$ this linear system \eqref{linearSystem} converges the fastest. Notice that this choice corresponds to a critically damped system (as opposed to overdamped or underdamped).

When the system is linear but multi-dimensional, $k$ can be assumed without loss of generality to be a symmetric matrix, and it can be immediately seen that there is no theoretical difficulty because one can diagonalize $k$ and choose $c$ diagonal wisely. Therefore, any numerical method that calculates the square root of a matrix could work here for getting $c$. There are many possible numerical approaches on square rooting matrices, for instance by preconditioning if the matrix has some special structure (which is usually the case in molecular systems), or as in \cite{Ha00} or \cite{Hi04}, but for consideration of conciseness the authors will not discuss this numerical topic.

\subsection{Temperature accelerated sampling: error bound}\label{Sectemaccsam1}
Denote by $\mu_N$ the distribution of $(q_N,p_N)$ using a cooling schedule $T(\cdot)$, by $\pi_{T(N)}$  the B-G distribution at  temperature $T(N)$, and by $h$ the integration time-step.

Assume that the Markov process of $(q_N,p_N)$ satisfies a uniform geometric ergodicity condition of the type
\begin{equation}\label{kadkhgsjdhgd}
    \| \mu_i-\pi_{T(i)} \|_{TV} \leq \rho_i \| \mu_{i-1} - \pi_{T(i)} \|_{TV}
\end{equation}
where $\pi_{T(i)}$ is the ergodic measure towards which the process converges if one step update from $(i-1)^{th}$ step to $i^{th}$ at temperature $T(i)$ is repeated, $\rho_i$ is the convergence rate, and statistical distance is measured in total variation norm, which is defined to be:
\begin{equation}
    \| \mu-\nu \|_{TV} = \sup_{A\in \mathcal{B}} |\mu(A)-\nu(A)|
\end{equation}
where $\mathcal{B}$ is the $\sigma$-algebra of measurable space.

By repetitive applications of triangle inequality, we derive from Equation \eqref{kadkhgsjdhgd} that:
\begin{equation}
    a_N \leq a_1 p_2 + \sum_{j=2}^N b_j p_j
\end{equation}
with $a_i=\|\mu_i-\pi_{T(i)}\|_{TV}$, $b_i=\|\pi_{T(i-1)}-\pi_{T(i)}\|_{TV}$, and $p_i=\prod_{k=i}^N \rho_k$.

We further assume that $0 < \rho_i \leq 1-\frac{h}{h_0}e^{-\frac{C_V}{T(i)}}$ for some constants $h_0$ (stable step length limit) and $C_V$ (elevation of potential energy).
Beyond transition state theory this assumption is motivated by  \cite{FW:98}, \cite{ScHu00}, \cite{DeZe98},  \cite{DiSt91}, \cite{BoOw:09} and \cite{MaStHi02}.

Using the assumption  $0 \leq T(j-1)-T(j) \ll T(j)$, we deduce a bound (function of the cooling schedule) on the sampling error:
\begin{equation}\label{bound}
\begin{split}
    \| \mu_N-\pi_N \|_{TV} \leq & \sum_{j=2}^N \left( \alpha_j \frac{T(j-1)-T(j)}{T(j)} \prod_{k=j}^N \left( 1-\frac{h}{h_0}e^{-\frac{C_V}{T(k)}} \right) + o\left(T(j-1)-T(j)\right) \right) \\&+ \prod_{k=2}^N \left( 1-\frac{h}{h_0}e^{-\frac{C_V}{T(k)}} \right)
    \end{split}
\end{equation}
where $\alpha_j =\mathbb{E}_{T(j)}[H]/T(j)$ ($\alpha_j=1$ for harmonic oscillators).

\subsection{Temperature accelerated sampling: cooling schedules}
\label{Sectemaccsam2}
 Naturally, one would like to minimize the error bound \eqref{bound} with respect to $T(n)$'s. This is however difficult because of nonlinearity. Instead, we consider the following subsets of cooling schedules (denote by $T_f$ the final temperature at which we want to sample the B-G, and by $N$ the number of steps we can afford to employ):

\paragraph{Inverse logarithmic cooling:}
\begin{equation}
    T(n)=T_f\frac{\log(N+1)}{\log(n+1)}
\end{equation}
This is the most popular schedule for optimization (\cite{GeGe84, Ha88}, for instance, have been frequently cited), but truncated at $T_f$ before $T\rightarrow 0$. Recall inverse logarithmic cooling is $T(n)=\frac{C}{\log(n+1)}$, and $C$ is fixed by requiring $T(N)=T_f$. When $N$ is fixed, there is no need to choose any parameter. This schedule will serve as our benchmark.

\paragraph{Shifted inverse logarithmic cooling:}
\begin{equation}
    T(n)=T_f+\frac{C}{\log(n+1)}
\end{equation}
where $C>0$ is the free parameter to be optimized. $T(N)$ is set to be $T_f$.

\paragraph{Exponential cooling:}
\begin{equation}
    T(n)=T_f e^{\tilde{C} (N-n)}=T_f C^{N-n}
\end{equation}
where $C=e^{\tilde{C}}>1$ is the free parameter to be optimized.

\paragraph{Shifted exponential cooling:}
\begin{equation}
    T(n)=T_f+\tilde{C}\cdot C^{-n}
\end{equation}
where $\tilde{C}>0$ and $C>1$ are free parameters. For ease on optimization, we chose $\tilde{C}=10^{-4} T_f C^N$ so that temperatures `smoothly' cool to $T_f$, and are left to optimize only one free parameter.

\paragraph{Linear cooling:}
\begin{equation}
    T(n)=\frac{n}{N}T_f+(1-\frac{n}{N})T_i
\end{equation}
where $T_i>T_f$ is the free parameter. This is used in \cite{Sz87} for optimization purposes. This seemingly too fast cooling schedule does give a small error bound in typical cases (see below).

\paragraph{Inverse linear cooling:}
\begin{equation}
    T(n)=1/\left( \frac{n}{N}\frac{1}{T_f}+(1-\frac{n}{N})\frac{1}{T_i} \right)
    \label{AnnealGLA}
\end{equation}
where $T_i>T_f$ is the free parameter. Instead of linearly interpolating the temperature, this linearly interpolates $\beta$ which is the inverse of temperature to ensure more steps at low temperatures.

\paragraph{Optimal error bound:}
We optimize error bounds \eqref{bound} for different total numbers of steps ($N$'s) with respect to the cooling schedules described above. As indicated by  Table \ref{differentNs}, the optimal schedule depends on the size on the total simulation time via
 $N$ (to be precise, the ratio between $Nh$ and  the mixing time of the original system). Unless $N$ is too small or too large, optimal inverse linear cooling produces a small error bound, optimal linear and exponential coolings have close performances as well, and all three optimal cooling schedules are similar. If the number of steps is too small, B-G will not be approximated well by any cooling schedule, and it is better to use the trivial schedule of constant temperature. If the number is instead too large (usually not the case of interest because accelerated sampling is desired), most types of cooling schedules will yield small errors, and surprisingly, shifted exponential cooling outperforms inverse logarithmic cooling, which is a popular cooling schedule for large $N$.

\begin{table}[h]
\footnotesize
\begin{tabular}{c|ccccccc}
\hline
N & Constant & Inverse log & Shifted inverse log & Exp & Shifted exp & Linear & Inverse linear \\
 & (no cooling) & (benchmark) & & & & & \\
\hline
200 & 0.896 & 1.304 & 0.950 & \textbf{0.896}$^{1}$ & \textbf{0.896}$^{1}$ & \textbf{0.896}$^{1}$ & \textbf{0.896}$^{1}$ \\
600 & 0.718 & 0.560 & 0.752 & 0.372$^{2}$ & 0.718 & \textbf{0.365}$^{2}$ & 0.368$^{2}$ \\
1000 & 0.575 &   0.325 &   0.597 &   0.266$^{3}$ &   0.346 &   0.267$^{3}$ &   \textbf{0.265}$^{3}$ \\
2000 & 0.331 &   \textbf{0.142} &   0.336 &   0.153$^{4}$ &   0.161 &   0.155$^{4}$ &   0.151$^{4}$ \\
5000 & 0.063 &   0.047 &   0.064 &   0.046$^{5}$ &   \textbf{0.028} &   0.047$^{5}$ &   0.046$^{5}$ \\
\hline
\end{tabular}
$^1$: Achieved by the limiting case of almost constant temperature \\
$^{2,3,4,5}$: Achieved by almost the same linear-alike optimizers within each row\\
\caption{\footnotesize Optimal error bound for different cooling schedules given total steps N. Within each row, bold indicates the minimum error bound. Different values of N are chosen to represent regimes of very small, small, medium, large, very large N's, in the sense of being compared to the total mixing steps which in this case renders the error bound 0.5 with a constant cooling and is $N\approx 1250$.}
\label{differentNs}
\end{table}

In these experiments, $T_f=20$, $C_V=150$, $h_0/h=1$, and $\alpha_j=1$. In this typical setting  $T_f/C_V$ is small and the B-G distribution is concentrated in potential wells, $h$ is close to $h_0$, and $\alpha_j \approx 1$. If the $T_f/C_V$ is large, however, the optimization  suggests not to anneal (result not shown). Optimization is done using MATLAB command fmincon.

\paragraph{Numerical validation on choices of cooling schedule:}

\begin{figure}[h]
\centering
\includegraphics[width=\textwidth]{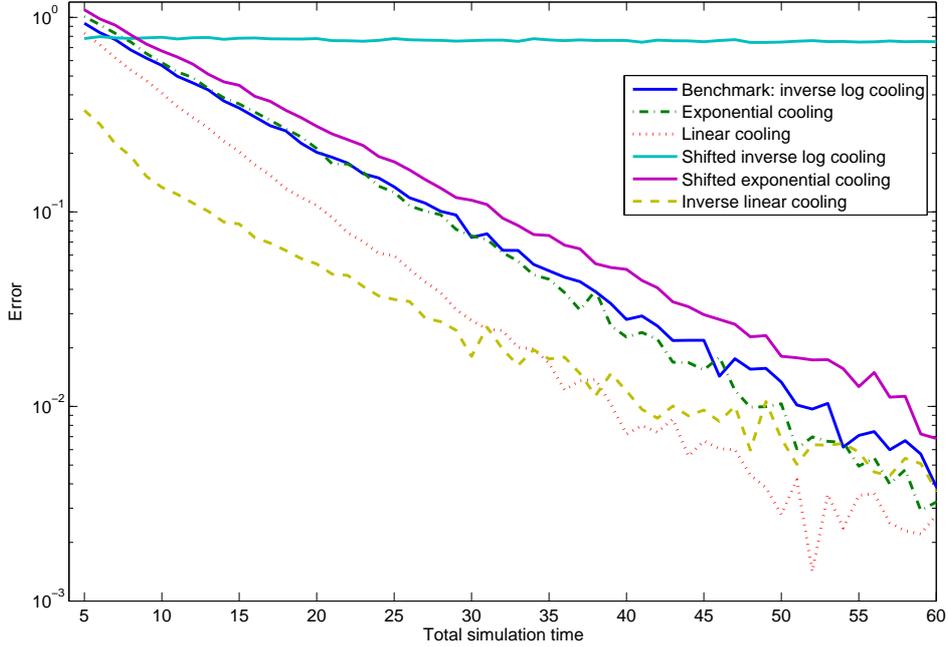}
\caption{\footnotesize Errors of representative cooling schedules as functions of total simulation time (hence of total simulation step $N$ too). Errors are calculated by $| \frac{1}{M}\sum_{i=1}^{M} q^i_N-\mathbb{E}q(\infty) |$, where $M=10000$ is the total number of independent trajectories, $q^i_N$ is the $N$th step position of the $i$th trajectory, $N\cdot h$ is the total simulation time and the step length $h=0.01$. $\mathbb{E}q(\infty)$ is well approximated by empirical average of an ensemble of $20000$ TuneGLA trajectories at total simulation time of $300$. Constants used in cooling schedules are: Shifted inverse log: $C=0.01T_f$, Exp: $C=1.5$, Shifted exp: $T(1)=2T_f$, Linear: $C=2T_f$, Inverse linear: $C=10T_f$. Basically all settings are the same as in Section \ref{exampleSection} except for total simulation time and cooling schedule used. Total simulation time is enumerated from 5 to 100 with an increment of 1.}
\label{DoubleLJstat}
\end{figure}

These cooling schedules have been  implemented on the example in Section \ref{exampleSection}. We did not optimize  cooling schedules with respect to free parameters  but used a heuristic/generic constant instead. Error on the empirical expectation of position has been investigated for each schedule in Figure \ref{DoubleLJstat}. The ranking of different types of schedules depends on total simulation time and agrees with theoretical prediction (except for large total simulation times which are dominated by numerical error accumulation).

\begin{figure}[h]
\centering
\includegraphics[width=\textwidth]{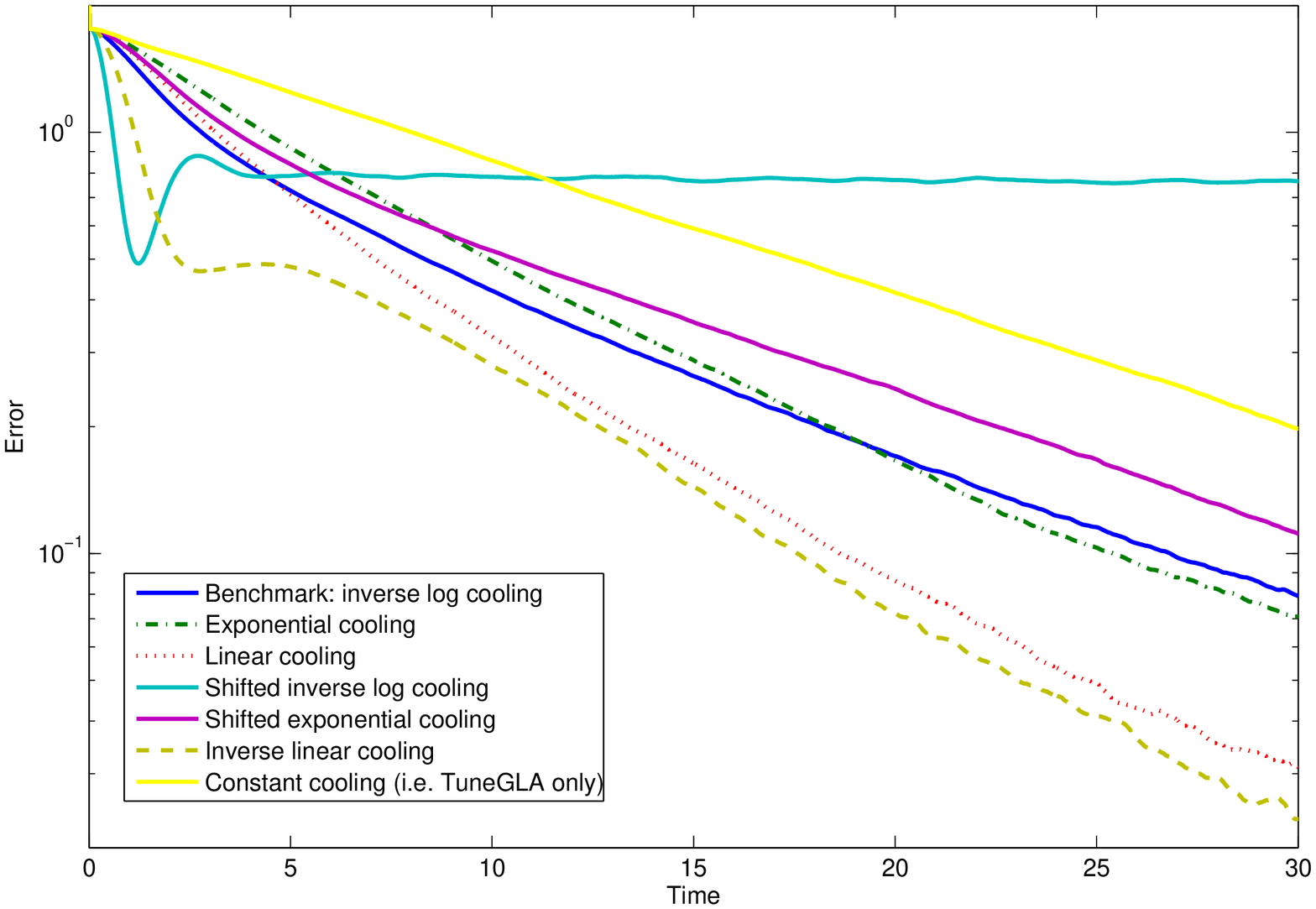}
\caption{\footnotesize Comparison of errors of TuneGLA with $c$ adaptively tuned and AnnealTuneGLA with different cooling schedules. Again, TuneGLA uses $\alpha=0.7$, total simulation time=30 is fixed, and all other settings are the same as in Figure \ref{DoubleLJstat} and \ref{GLAcomparison} too.}
\label{coolingComparison}
\end{figure}

In addition to Figure \ref{DoubleLJstat} and the above discussion that compare cooling schedules for different total simulation times, we fix total time and show time dependent errors of different schedules in Figure \ref{coolingComparison}. Here total simulation time is $30$ and we are in the medium $N$ regime. Inverse linear cooling indeed has better performances, followed closely by linear cooling, both consistent with the theoretical analysis. Rigorously speaking one should compare cooling schedules only towards the end of the simulation, because different cooling schedules are at different temperatures in the middle of the simulation; however, the superiority of inverse linear cooling is in fact exhibited throughout the simulation.

These numerical experiments and theoretical bounds  indicate that inverse linear cooling is ranked at the top.
It is worth pointing out that although annealing accelerates convergence significantly, one has to choose a priori parameters (in most of our cases, total simulation step $N$ and constant $C$ or $T_i$). This issue usually needs a case-by-case investigation, but  $C_V$ (if known)  could be used in conjunction with the error bound to determine $N$ and $C$.

\end{document}